%% file: ariane.tex
\documentclass[journal]{IEEEtran}

\usepackage{graphicx}
\usepackage{booktabs}
\usepackage{tabularx}
\usepackage{multirow}
\usepackage{multicol}
\usepackage{color}
\usepackage[binary-units]{siunitx}
\usepackage[acronym]{glossaries}
\usepackage{blindtext}
\usepackage{pgfplots}
\usepackage[printwatermark]{xwatermark}
\usepackage{paralist}

\pgfplotsset{compat=1.14}

\usepackage{threeparttable}

\graphicspath{{figures/}}
\newcommand{\Secref}[1]{Section~\ref{#1}}

\renewcommand{\arraystretch}{1.2}

\DeclareSIUnit\op{op}

\DeclareSIUnit\OP{op}
\DeclareSIUnit\OPs{\OP\per\s}
\DeclareSIUnit\OPsW{\OP\per\s\per\watt}
\DeclareSIUnit\GOPs{\giga\OPs}
\DeclareSIUnit\GOPsW{\giga\OPsW}
\DeclareSIUnit\TOPs{\tera\OPs}
\DeclareSIUnit\TOPsW{\tera\OPsW}

\DeclareSIUnit\FLOP{flop}
\DeclareSIUnit\FLOPs{\FLOP\per\s}
\DeclareSIUnit\FLOPsW{\FLOP\per\s\per\watt}
\DeclareSIUnit\GFLOPs{\giga\FLOPs}
\DeclareSIUnit\GFLOPsW{\giga\FLOPsW}
\DeclareSIUnit\TFLOPs{\tera\FLOPs}
\DeclareSIUnit\TFLOPsW{\tera\FLOPsW}

\begin{document}

\title{The Cost of Application-Class Processing: Energy and Performance Analysis of a Linux-ready 1.7GHz 64bit RISC-V Core in 22nm FDSOI Technology}
\author{Florian~Zaruba,~\IEEEmembership{Student Member,~IEEE,}
        and~Luca~Benini,~\IEEEmembership{Fellow,~IEEE}

\thanks{The authors are with the Integrated Systems Laboratory of ETH Zurich,
Zurich, Switzerland (e-mail: zarubaf@iis.ee.ethz.ch; lbenini@iis.ee.ethz.ch).}
}

\newacronym{bht}{BHT}{Branch History Table}
\newacronym{btb}{BTB}{Branch Target Buffer}
\newacronym{ras}{RAS}{Return Address Stack}
\newacronym{soc}{SoC}{System on Chip}
\newacronym{axi}{AXI}{Advanced eXtensible Interface}
\newacronym{risc}{RISC}{Reduced Instruction Set Computer}
\newacronym{csr}{CSR}{Control and Status Registers}
\newacronym{rob}{ROB}{Re-order Buffer}
\newacronym{lsu}{LSU}{load/store unit}
\newacronym{fpu}{FPU}{Floating Point Unit}
\newacronym{alu}{ALU}{Arithmetic Logic Unit}
\newacronym{ptw}{PTW}{Page Table Walker}
\newacronym{tlb}{TLB}{Translation Lookaside Buffer}
\newacronym{plru}{LRU}{Pseudo Least Recently Used}
\newacronym{os}{OS}{Operating System}
\newacronym{isa}{ISA}{Instruction Set Architecture}
\newacronym{ipc}{IPC}{Instructions per Cycle}
\newacronym{mmu}{MMU}{Memory Management Unit}
\newacronym{smp}{SMP}{Symmetric Multiprocessing}
\newacronym{ge}{GE}{Gate Equivalent}
\newacronym{waw}{WAW}{Write after Write}
\newacronym{pc}{PC}{Program Counter}
\newacronym{simd}{SIMD}{Single Instruction Multiple Data}

\markboth{
  IEEE Transaction on Very Large Scale Integration (VLSI) Systems,~Vol.~{(vol)}, No.~{(no)}, {(month)}~{(year)}
}{%
  \ifCLASSOPTIONpeerreview\else Zaruba \MakeLowercase{\textit{et al.}}: \fi
  The cost of Application-Class Processing: Energy and performance analysis of a linux-ready 1.7GHz 64bit RISC-V Core in 22nm FDSOI technology
}

\maketitle
\IEEEpubid{\begin{minipage}{\textwidth}\ \\[12pt] \centering
  \copyright 2019  IEEE.  Personal use of this material is permitted.  Permission from IEEE must be obtained for all other uses, in any current or future media, including reprinting/republishing this material for advertising or promotional purposes, creating new collective works, for resale or redistribution to servers or lists, or reuse of any copyrighted component of this work in other works.
\end{minipage}}

\begin{abstract}
The open-source RISC-V ISA \cite{waterman2014risc} is gaining traction, both in industry and academia. The ISA is designed to scale from micro-controllers to server-class processors. Furthermore, openness promotes the availability of various open-source and commercial implementations. Our main contribution in this work is a thorough power, performance, and efficiency analysis of the RISC-V ISA targeting baseline ``application class'' functionality, i.e. supporting the Linux OS and its application environment based on our open-source single-issue in-order implementation of the 64 bit ISA variant (RV64GC) called Ariane. Our analysis is based on a detailed power and efficiency analysis of the RISC-V ISA extracted from silicon measurements and calibrated simulation of an Ariane instance (RV64IMC) taped-out in GlobalFoundries 22\,FDX technology. Ariane runs at up to 1.7\,GHz and achieves up to 40\,Gop/sW peak efficiency. We give insight into the interplay between functionality required for application-class execution (e.g. virtual memory, caches, multiple modes of privileged operation) and energy cost. Our analysis indicates that ISA heterogeneity and simpler cores with a few critical instruction extensions (e.g. packed SIMD) can significantly boost a RISC-V core's compute energy efficiency.
\end{abstract}


\IEEEpeerreviewmaketitle

\input{sec_intro}
\input{sec_arch}
\input{sec_impl}
\input{sec_results}
\input{sec_relwork}
\input{sec_conc}

\section*{Acknowledgments}
The authors would like to thank Michael Schaffner and Fabian Schuiki for comments that greatly improved the manuscript.

This work has received funding from the European Union’s Horizon 2020 research and innovation program under grant agreement No 732631, project ``OPRECOMP''.


\ifCLASSOPTIONcaptionsoff
  \newpage
\fi

\bibliographystyle{IEEEtran}
\bibliography{IEEEabrv,ref}

\begin{IEEEbiography}[{\includegraphics[width=1in,height=1.25in,clip,keepaspectratio]{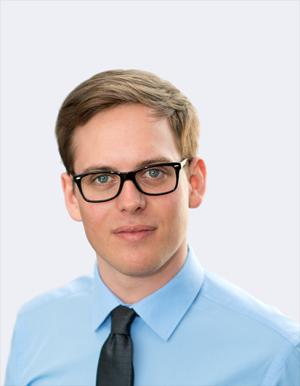}}]{Florian Zaruba}
  received his BSc degree from TU Wien in 2014 and his MSc from the Swiss Federal Institute of Technology Zurich in 2017. He is currently pursuing a PhD degree at the Integrated Systems Laboratory. His research interests include design of very large scale integration circuits and high performance computer architectures.
\end{IEEEbiography}

\begin{IEEEbiography}[{\includegraphics[width=1in,height=1.25in,clip,keepaspectratio]{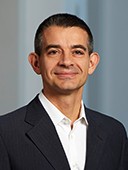}}]{Luca Benini}
    holds the chair of digital Circuits and systems at ETHZ and is Full Professor at the Universita di Bologna. Dr. Benini's research interests are in energy-efficient computing systems design, from embedded to high-performance. He has published more than 900 peer-reviewed papers and five books. He is a Fellow of the ACM and a member of the Academia Europaea. He is the recipient of the 2016 IEEE CAS Mac Van Valkenburg award.
\end{IEEEbiography}

\vfill

\end{document}

%% file: sec_intro.tex
\section{Introduction}
\IEEEpubidadjcol

\IEEEPARstart{T}{he} relatively new open \gls{isa} RISC-V has already seen wide-spread adoption in industry and academia~\cite{greenwaves,sifive,westerndigital,patsidis2018low}. It is based on \gls{risc} principles and heavily relies on standard and non-standard extensions to tailor the ISA without cluttering the instruction set base. Furthermore, the base ISA is split into privileged and non-privileged instructions. While the non-privileged ISA governs the base instruction set, the privileged ISA includes different levels of hardware support needed to run an \gls{os}. Common and standardized RISC-V extensions are integer (I), multiply/divide support (M), atomic memory operations (A), and single (F) and double (D) precision IEEE floating point support. Together they form the general purpose processing extension (G). Furthermore, the RISC-V ISA supports variable length instructions. Currently, it only defines 32 and 16 bit compressed instructions (C).

The instruction set was designed to scale from micro-controllers to server-class platforms. While there already exist a plethora of open micro-controller cores  \cite{gautschi2017near,wolf2018picorv32}, there are fewer cores available in the higher, Linux-capable, performance range, mostly due to increased design and verification costs. CPUs which offer support for UNIX-like \gls{os}es are usually called application-class processors. The hardware overhead to efficiently support \gls{os}es like Linux is significant: To enable fast address translation a \gls{tlb} and a \gls{ptw} are needed. A Linux-like \gls{os} needs at least a few dozen \SI{}{\mega\byte} of main memory. In most of the cases, this memory will be off-chip making it inefficient to be accessed constantly and requiring some sort of caching mechanism. Cache look-up and address translation are often on the critical path in modern CPU designs as accessing memory is slow. Still, keeping the operating frequency high is of importance since \gls{os}es contain large portions of highly serial code. This requires further pipelining and more sophisticated hardware techniques to hide the increased pipeline-depth such as scoreboarding, branch-prediction and more elaborate out-of-order techniques \cite{hennessy2011computer, patterson2013computer}. This increases both static and dynamic power. Furthermore, the advent of \gls{smp} support in \gls{os}es made it necessary to provide efficient and fast atomic shared memory primitives in the \gls{isa}. RISC-V provides this as part of the A-extension in the form of load-reserve and store-conditional as well as atomic fetch-and-op instructions which can perform operations like integer arithmetic, swapping and bit-wise operations close to memory.

Nevertheless, there are significant gains in having support for a full-blown \gls{os}: The \gls{os} eases programmability by providing a standardized interface to user programs, memory-management, isolation between user programs, and a vast amount of libraries, drivers, and user programs. Furthermore full-featured \gls{os}es  (e.g. Sel4~\cite{klein2009sel4}) provide an additional guaranteed layer of security.

\IEEEpubidadjcol

Energy-efficiency is becoming the paramount design goal for the next generation architectures \cite{kiamehr2017temperature,hwang2016comparative}. Due to its modularity, the RISC-V instruction set offers a wide variety of microarchitectural design choices, ranging from low-cost microcontroller to high performance, out-of-order, server-class CPUs supporting Linux \cite{Celio:EECS-2017-157}. This modularity makes the \gls{isa} suitable for a more thorough analysis of different architectural features and their impact on energy efficiency.

Our work aims at giving insight on the energy cost and design trade-offs involved in designing a RISC-V core with support for a full-featured \gls{os}. For a thorough analysis, we designed a competitive 64\,bit, 6-stage, in-order application-class core which has been manufactured in \textsc{Globalfoundries 22\,FDX} technology. We performed extensive silicon characterization and a detailed, per-unit efficiency analysis based on silicon-calibrated post-layout simulations. The core runs at \SI{1.7}{\giga\hertz} and achieves an efficiency of up to \SI{40}{\GOPsW}.

In particular, our main contributions in this work are:
\begin{itemize}
    \item Implementation of an in-order, single-issue, 64\,bit application-class processor called Ariane. Ariane has been open-sourced on GitHub.\footnote{https://github.com/pulp-platform/ariane}
    \item Silicon integration of Ariane in a state-of-the-art 22\,nm SOI process.
    \item Exploration of trade-offs in performance and efficiency based on silicon measurements.
    \item Thorough analysis of the RISC-V ISA and Ariane's microarchitecture based on measurements and silicon-calibrated post-layout simulations.
\end{itemize}

We explore Ariane's microarchitecture in detail in the next section. In Sec.~\ref{sec:impl} we touch upon physical design of a particular Ariane instance. Finally Sec.~\ref{sec:results} contains a detailed power, performance and efficiency analysis.

%% file: sec_arch.tex
\section{Architecture} 
\label{sec:arch}

\begin{figure*}
  \centering
  \includegraphics[width=\linewidth]{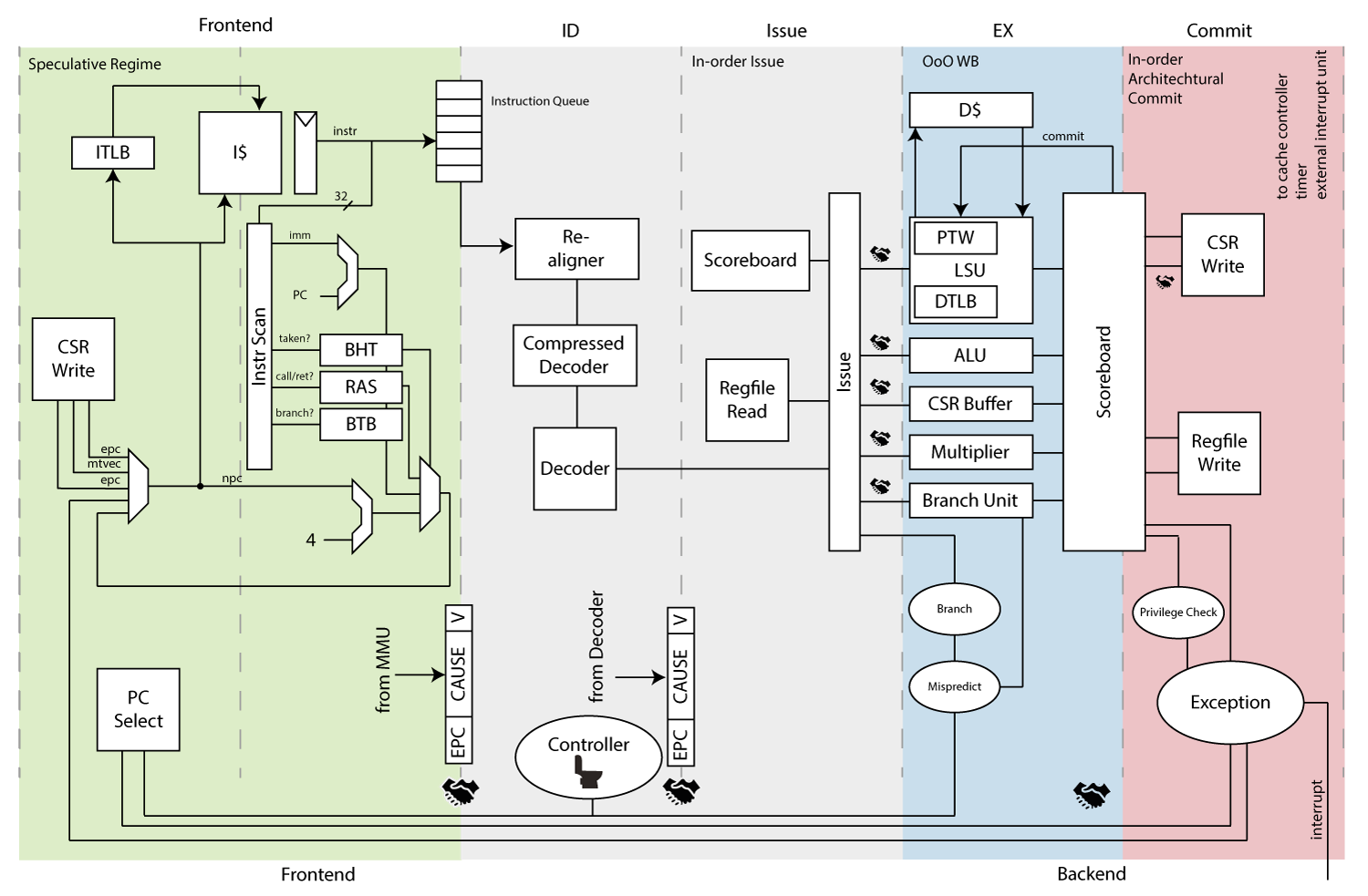}
  \caption{\label{fig:ariane_bd}Blockdiagram of Ariane; Depicted are the six separate stages which exchange data via handshaking. The scoreboard logically wraps the execute stage (cf.~\ref{subsec:scoreboard}) and provides an interface for the issue and commit stages. For a detailed description of each pipeline stage and functional unit see~\Secref{sec:arch}.}
\end{figure*}

Ariane is a 64\,bit, single-issue, in-order RISC-V core. It has support for hardware multiply/divide, atomic memory operations as well as an IEEE compliant \gls{fpu}. Furthermore, it supports the compressed instruction set extension as well as the full privileged instruction set extension. It implements a 39\,bit, page-based virtual memory scheme (SV39). The primary design goal of the microarchitecture was to reduce critical path length while keeping \gls{ipc} losses moderate. The target logic depth was chosen to be below 30 NAND \glspl{ge} which is just a factor of two higher than state-of-the-art, highly tuned, server-class, out-of-order processors \cite{bowhill2016xeon}. To achieve desired performance goals a synthesis-driven design approach lead to a 6-stage pipelined design.
To reduce the penalty of branches, the microarchitecture features a branch predictor. A high-level block diagram is depicted in Figure~\ref{fig:ariane_bd}. In the following, we give a stage-by-stage description of the complete microarchitecture.
\vspace{4mm}
\begin{table}[!t]
    \renewcommand{\arraystretch}{1.3}
        \caption{RISC-V Static Binary Analysis (riscv64-unknown-linux-gnu-gcc 7.2.0)}
        \label{tab:static_binary_analysis}
        \centering
        \begin{tabular}{@{}llll@{}}
        \toprule
        \bfseries Benchmark~\cite{coremarkpro} & \bfseries Compr. Ratio & \bfseries Branches [\%] & \bfseries Calls [\%] \\
        \midrule
        cjpeg-rose7 & 0.71 & 34.78 & 6.71 \\
        dhrystone & 0.72 & 35.36 & 6.78 \\
        linear-alg-mid-100x100-sp & 0.72 & 33.65 & 6.17 \\
        loops-all-mid-10k-sp & 0.72 & 33.78 & 6.22 \\
        nnet & 0.72 & 33.75 & 6.17 \\ 
        parser-125k & 0.72 & 33.73 & 6.43 \\
        radix2-big-64k & 0.85 & 22.77 & 1.99 \\
        sha & 0.72 & 33.25 & 6.01 \\
        zip & 0.72 & 33.48 & 6.21 \\
    \bottomrule
    \end{tabular}
\end{table}
\begin{enumerate}
    \item \textbf{PC Generation}  is responsible for selecting the next \gls{pc}. This can either come from \gls{csr} when returning from an exception, the debug interface, a mispredicted branch, or a consecutive fetch.
    \item \textbf{Instruction Fetch} contains the instruction cache, the fetch logic and the pre-decode logic which guides the branch-prediction of the \gls{pc} stage. 
    \paragraph{Address Translation} 
    The instruction cache is virtually indexed and physically tagged and fully parametrizable. The \gls{pc} calculated by the previous stage is split into page-offset (lower 12 bit) and virtual page number (bit 12 up to 39). The page-offset is used to index into the instruction cache in the first cycle while the virtual page number is simultaneously used for address translation through the instruction \gls{tlb}. In case of a \gls{tlb} miss the cache pipeline is stalled until the translation is valid. 
    \paragraph{Cache Pipelining}
    Data coming from the instruction cache's data arrays is registered before being pre-decoded to mitigate the effects of the long propagation delay of the memory macros in the cache. This has the side-effect that even on a correct control flow prediction we will lose a cycle as we can not calculate the next \gls{pc} in the same cycle we receive the data from the instruction cache. With the additional compressed instruction set extension, this is usually not a problem as (with a fetch width of 32 bit) we are fetching 1.5 instructions on average, and approximately 70\% of all instructions are compressed. Furthermore approximately 35\% of the instructions are branches (see Table~\ref{tab:static_binary_analysis}). This means we can easily tolerate a single cycle delay on branch-prediction (caused by the additional register stage after the instruction cache) and still generate a consecutive instruction stream for the processor's back-end.
    \paragraph{Frontend}
    Together with the \gls{pc} stage, the instruction fetch stage forms the processor's frontend. The frontend is fully decoupled from the back-end, which is in charge of executing the instruction stream. The decoupling is implemented as a FIFO of configurable depth. Instructions are stored in compressed form in the queue minimizing the number of flip-flops necessary for the instruction FIFO. Mis-predicted control flow instructions are resolved during the execute stage in a specialized branch unit~\cite{gonzalez2010processor}. 
    \item \textbf{Instruction Decode} re-aligns potentially unaligned instructions, de-compresses them and decodes them. Decoded instructions are then put into an issue queue in the issue stage.
    \item \textbf{Issue Stage} contains the issue queue, a scoreboard, and a small \gls{rob}. Once all operands are ready the instruction is issued to the execute stage. Dependencies are tracked in the scoreboard and operands are forwarded from the \gls{rob} if necessary.
    \item \textbf{Execute Stage} houses all functional units. Every functional unit is handshaked and readiness is taken into account during instruction issue. Furthermore, we distinguish between fixed and variable latency units. Fixed latency are the integer \gls{alu}, multiplier/divider and \gls{csr} handling. The only variable latency units are currently the \gls{fpu} and the \gls{lsu}. Instructions can retire out-of-order from the functional units. Write-back conflicts are resolved through the \gls{rob}.
    \item \textbf{Commit Stage} reads from the \gls{rob} and commits all instructions in program order. Stores and atomic memory operations are held in a store buffer until the commit stage confirms their architectural commit. Finally, the register file is updated by the retiring instruction. To avoid artificial starvation because of a full \gls{rob} the commit stage can commit two instructions per cycle.
\end{enumerate}

Next, we describe the main units of the microarchitecture, summarizing key features. 

\subsection{Branch Prediction}


As the pipeline depth of processors increases the cost for mis-predicted branches rises significantly. Mis-prediction can occur on the jump target address (the jump address is determined by a register value) as well as on a mis-predicted branch outcome. On mis-prediction the frontend as well as the decode and issue stages need to be flushed, which introduces at least a five-cycle latency in the pipeline, and even more on \gls{tlb} or instruction cache misses. 

To mitigate the negative impact of control flow stalls on \gls{ipc}, Ariane contains three different means of predicting the next \gls{pc}, namely a \gls{bht}, \gls{btb}, and \gls{ras}. To facilitate branch-prediction Ariane does a light pre-decoding in its fetch interface to detect branches and jumps. It furthermore needs to re-align instruction fetches as interleaved compressed instructions (16 bit instructions) can offset regular (32 bit) instructions effectively making it possible for an instruction to wrap the 32\,bit fetch boundary.

A classic two bit saturation counter \gls{bht} is used for predicting on the branch outcome. Branches in RISC-V can only jump relative to the \gls{pc} which makes it possible to redirect control flow immediately. If there is no valid prediction available static prediction is being used as a fall-back strategy. Static prediction in RISC-V is defined as backward jumps (negative immediate) always taken and forward jumps (positive immediate) never taken, hence they can be decided very efficiently by looking at a single bit in the immediate field. Furthermore, the ISA provides \gls{pc}-relative and absolutely addressed control flow changes. \gls{pc}-relative unconditional jumps can be resolved as soon as the instruction is being fetched. Register-absolute jumps can either be predicted using the \gls{btb} or the \gls{ras} depending on whether the jump is used as a function call or not. 

\subsection{Virtual Memory}

To support an operating system Ariane features full hardware support for address translation via a \gls{mmu}. It has separate, configurable data and instruction \gls{tlb}s. The \gls{tlb}s are fully set-associative, flip-flop based, standard-cell memories. On each instruction and data access, they are checked for a valid address translation. If none exists, Ariane's hardware \gls{ptw} queries the main memory for a valid address translation. The replacement strategy of \gls{tlb} entries is \gls{plru}~\cite{bhattacharjee2017architectural}. 

\subsection{Exception Handling}

Exceptions can occur throughout the pipeline and are hence linked to a particular instruction. The first exception can occur during instruction fetch when the \gls{ptw} detects an illegal \gls{tlb} entry. During decode, illegal instruction exceptions can occur while the \gls{lsu} can also fault on address translation or trigger an illegal access exception. As soon as an exception has occurred the corresponding instruction is marked and auxiliary information is saved. When the excepting instruction finally retires the commit stage redirects the instruction frontend to the exception handler.

Interrupts are asynchronous exceptions which are synchronized to a particular instruction. This results in the commit stage waiting for a valid instruction to retire, in order to take an external interrupt and associating an exception with it. Atomic memory operations must not be interrupted, which simply translates to not taking an interrupt when we retire an atomic instruction. The same holds true for atomic \gls{csr} instructions which can alter the hart's (HARdware Thread) architectural state.

\subsection{Privileged Extensions}
In addition to virtual memory, the privileged specification defines more \gls{csr}s which govern the execution mode of the hart. The base supervisor \gls{isa} defines an additional interrupt stack for supervisor mode interrupts as well as a restricted view of machine mode \gls{csr}s. Access to these registers is restricted to the same or a higher privilege level. 

CSR accesses are executed in the commit stage and are never done speculatively. Furthermore, a \gls{csr} access can have side-effects on subsequent instructions which are already in the pipeline and have been speculatively executed e.g. altering the address translation infrastructure. This makes it necessary to completely flush the pipeline on such accesses.

In addition to the base \gls{isa} the privileged ISA defines a handful more instructions ranging from power hints (sleep and wait for interrupt) to changing privilege levels (call to the supervising environment as well as returning). As those instructions alter the \gls{csr} state as well as the privilege level they are only executed non-speculatively in the commit stage. 

The RISC-V \gls{isa} defines separate memory streams for instruction, data, and address translation, all of which need to be separately synchronized with special memory ordering instructions (fences). For caches, this means that they are either coherent or need to be entirely flushed. As the \gls{tlb}s in Ariane are designed with flip-flops they can be efficiently flushed in a single cycle. 

\subsection{Register Files}

The core provides two physically different register files for floating-point and integer registers. We provide the choice of either a latch-based or flip-flop-based implementation. The advantage of the latch-based approach is that it is approximately half the area of the flip-flop version. A known duty cycle is of importance when using a latch-based register file as capturing is happening on the falling edge of the clock. Therefore (high-speed) clock generators need to take care of a balanced and low jitter duty-cycle.

\subsection{Scoreboard/Reorder Buffer}
\label{subsec:scoreboard}
The scoreboard, including the \gls{rob}, is implemented as a circular buffer which logically sits between the issue and execute stage and contains:
\begin{itemize}
    \item Issued, decoded, in-flight instructions which are currently being executed in the execute stage. Source and destination registers are tracked and checked by the issue stage to track data hazards. As soon as a new instruction is issued, it is registered within the scoreboard.
    \item Speculative results written back by the various functional units. As the destination register of each instruction is known, results are forwarded to the issue stage when necessary. The commit stage reads finished instructions and retires them, therefore making room for new instructions to enter the scoreboard.
\end{itemize}

\gls{waw} hazards in the scoreboard are resolved through a light-weight re-naming scheme which increases the logical register address space by one bit. Each issued instruction toggles the MSB of its destination register address and subsequent read addresses are re-named to read from the latest register address.

\subsection{Functional Units}

Ariane contains 6 functional units:

\begin{enumerate}
    \item \textbf{\gls{alu}}: Covers most of the RISC-V base ISA, including branch target calculation.
    \item \textbf{\gls{lsu}}: Manages integer and floating-point load/stores as well as atomic memory operations. The \gls{lsu} interfaces to the data cache using three master interfaces. One dedicated for the \gls{ptw}, one for the load unit while the last one is allocated to the store unit. The data cache is a parameterizable write-back cache which supports hit-under-miss functionality on the different master ports. In addition, the store unit contains a variable-size store buffer to hide the store latency of the data cache. Ideally, the store-buffer is sized so that context store routines commonly found in \gls{os} code can be retired with an \gls{ipc} of 1.
    \item \textbf{\gls{fpu}}: Ariane contains an IEEE compliant floating-point unit with custom trans-precision extensions~\cite{tagliavini2018transprecision,mach2018transprecision}
    \item The \textbf{branch unit} is an extension to the \gls{alu} which handles branch-prediction and branch-correction.
    \item \textbf{\gls{csr}}: RISC-V mandates atomic operations on its \gls{csr}, as they need to operate on the most up-to-date value Ariane defers reading or writing until the instruction is committed in the commit stage. The corresponding write data is buffered in this functional unit and read again when the instruction is retiring. 
    \item \textbf{Multiplier/Divider}: This functional unit houses the necessary hardware support for the M-extension. The multiplier is a fully pipelined 2-stage multiplier. We rely on re-timing to move the pipeline register into the combinational logic during synthesis. The divider is a bit-serial divider with input preparation. Depending on the operand values division can take from 2 to 64 cycles. 
\end{enumerate}

\subsection{Caches}

Both data and instruction caches are virtually indexed and physically tagged. Set-associativity and cache-line size of both caches can be adapted to meet core area constraints and timing. As even fast cache memories are relatively slow compared to logic, the instruction and data cache both have an additional pipeline stage on their outputs. This allows for relatively easy path balancing by means of de-skewing (i.e. adjusting the memories clock to arrive earlier or later than the surrounding logic). 

\subsection{Memory and Control Interfaces}

The core contains a single \gls{axi} 5 master port as well as four interrupt sources: 
\begin{compactenum}
    \item Machine External Interrupts: Machine-mode platform interrupts e.g. UART, SPI, etc.
    \item Supervisor External Interrupts: Supervisor-mode platform interrupts (i.e. the \gls{os} is in full control)
    \item Machine Timer Interrupt: Platform timer (part of the RISC-V privileged specification). Used by the \gls{os} for time-keeping and scheduling.
    \item Machine Software Interrupt: Inter processor interrupts
\end{compactenum}

The master port is arbitrated between instruction fetch, data cache re-fill and write-back as well as cache bypass (i.e. un-cached) accesses.

\subsection{Debug Interface}

Ariane contains a RISC-V compliant debug interface~\cite{riscvdebug}. For the implementation of the debug interface an execution-based approach was chosen to keep the debug infrastructure minimally invasive on the microarchitecture and therefore improving on efficiency and critical path length (e.g. no multiplexers on \gls{csr} or general purpose registers). The core uses its existing capability to execute instructions to facilitate debugging by fetching instructions from a debug RAM. To put the core into debug mode an interrupt-like signal is asserted by the debug controller and the core will jump to the base address of the debug RAM. Only one additional instruction (dret) is required to return from debug mode and continue execution. Communication with the external debugger is done through a debug module (DM) peripheral situated in the peripheral clock and power domain.

\subsection{Tracing and Performance Counters}

We support extensive tracing in RTL simulation. All register and memory access are traced together with their (physical and virtual) addresses and current values. Currently, we do not support \gls{pc} tracing in hardware.

Performance counters are mapped into the \gls{csr} address space. In particular, we support counting the following events: cycle count, instructions-retired, L1 instruction/data cache miss, instruction/data \gls{tlb} miss, load/store instruction counter, exception counter and various branch-prediction metrics.

\subsection{Concluding Remarks}

The design of a fast processor with a reasonably high \gls{ipc} poses some interesting challenges. Significant complexity revolves around the L1 memory interface which ideally should be large, low latency and fast (running at the speed of the core's logic). The introduction of virtual memory adds to the already existing complexity. Large portions of the design are built around the idea to hide (especially the load) memory latency by cleverly scheduling other instructions and trying to do as much useful work in each clock cycle. Furthermore, the introduction of more privileged architectural state in the form of virtual to physical address translation as well as additional \gls{csr}s results in an increased area and (leakage) power. These cost factors are analyzed in details in Sec. \ref{sec:results}.


%% file: sec_impl.tex
\section{Implementation}
\label{sec:impl}

\begin{table}[!t]
    \renewcommand{\arraystretch}{1.3}
        \caption{Architectural Design Choices for Silicon Implementation}
        \label{tab:arch_choices}
        \centering
        \begin{tabular}{@{}ll@{}}
        \hline
        \bfseries Parameter & \bfseries Chosen\\
        \hline
        \gls{bht} & 8 \\
        \gls{btb} & 8 \\
        \gls{rob} Entries & 8 \\
        Fetch latency & 1 \\
        L1 I-cache (4-way) size & \SI{16}{\kilo\byte} \\
        L1 D-cache (8-way) size & \SI{32}{\kilo\byte} \\
        L1 D-cache latency & 3 \\
        Integer \gls{alu} latency & 1 \\
        Register File & 31x64 flip-flops \\
    \hline
    \end{tabular}
\end{table}

\begin{figure}
    \includegraphics[width=\linewidth]{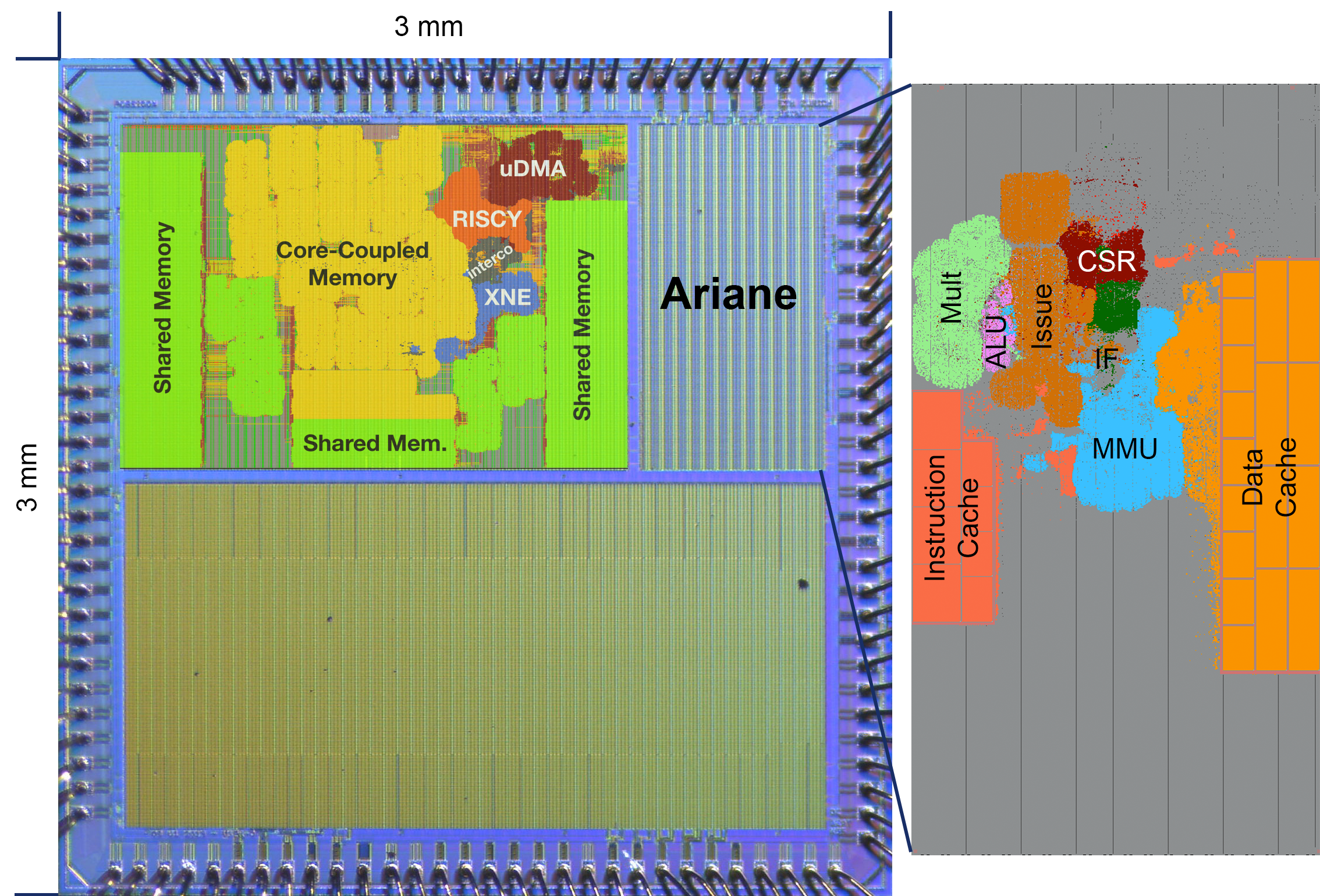}
    \caption{\label{fig:floorplan}Ariane has been implemented in \textsc{Globalfoundries}\SI{22}{\nano\meter} on a \SI{3x3}{\milli\meter} die. The SoC has been separately hardened as a macro and is depicted on the left while a more detailed floorplan of Ariane can be seen on the right.}
\end{figure}

Ariane has been taped-out in \textsc{Globalfoundries 22\,FDX}. The coreplex has been hardened as a macro and uses a shared \gls{soc} infrastructure for off-chip communication \cite{schiavone2018quentin}. The Ariane coreplex can communicate with the SoC via a full-duplex 64\,bit data and address \gls{axi} interconnect. The SoC contains \SI{520}{\kilo\byte} of on-chip scratchpad memory and an extensive set of peripherals such as HyperRAM, SPI, UART and I2C. The core is separately clocked by a dedicated frequency locked loop (FLL) in the SoC domain.

The Ariane coreplex can be separately supplied and powered down. Furthermore, the logic cells can be forward body biased (FBB) to increase speed at the expense of leakage power. The taped-out instance of Ariane contains \SI{16}{\kilo\byte} of instruction cache with a set-associativity of four. The data cache is \SI{32}{\kilo\byte} in size with a set-associativity of eight. The instruction and data \gls{tlb} are each 16 entries in size. The architectural parameter settings are listed in Table~\ref{tab:arch_choices}.

The design has been synthesized using Synopsys Design Compiler 2016.12. The back-end design flow has been carried out using Cadence Innovus-16.10.000. We use an eight track, multi-threshold, multi-channel, standard cell library with nominal voltage at \SI{0.8}{\volt} from \textsc{Invecas}. For the cache memories, we use a high-performance, single-port register file generator provided by \textsc{Invecas}. The design has been signed-off at \SI{902}{\mega\hertz} at \SI{0.72}{\volt}, \SI{125}{\celsius}, SSG. The floorplan of the chip is depicted in Figure~\ref{fig:floorplan}. The final netlist contains 75.34\% LVT (low voltage threshold) and 24.66\% SLVT (super low voltage threshold) cells.

\subsection{Physical Design}

To achieve higher clock speeds several optimizations have been applied during physical design: Useful-skew has been used for placement, clock-tree synthesis and routing to balance the critical paths from and to the memories. Clock-shielding was employed to mitigate the effects of cross-talk on the clock tree. Furthermore, we have placed decap cells close to clocktree buffers. Together with a dense power grid, this mitigates the effects of IR drop. 

The most critical paths with a gate delay of 30 are around the data caches as the high set-associativity of eight significantly drains valuable routing resources close to the memory macros. Therefore, special routing channels have been provisioned in the floorplan between the memory macros. Furthermore, due to the high hold-times of the memory macros, special attention has been paid to hold-time fixing on those paths.

A multi-mode, multi-corner (MMMC)\footnote{Total of 21 corners: functional-mode, worst RC, best RC, SSG, TT, FFG, \SIrange{0.72}{0.88}{\volt}, \SIrange{-40}{125}{\celsius}} with AOCV views (Advanced On Chip Variations) approach has been used for the entire back-end flow to reduce the margin we need to provision on clock frequency. 



%% file: sec_results.tex
\section{Results} 
\label{sec:results}

\begin{table*}[!t]
    \renewcommand{\arraystretch}{1.3}
        \caption{Energy per Operation Class [\SI{}{\pico\joule}], Leakage [\SI{}{\milli\watt}]}
        \label{tab:energy_per_op}
        \centering
        \begin{tabular}{@{}lrrrcrrrrrrrrrrrrr@{}}
        \toprule
        \bfseries Instr. Class & \bfseries PC & \multicolumn{2}{c}{\bfseries IF Stage} & \multicolumn{2}{c}{\bfseries ID Stage} & \bfseries Issue & \multicolumn{6}{c}{\bfseries EX Stage} & \bfseries WB & \bfseries CSR & \bfseries CTS & \bfseries Rest & \bfseries Tot \\
        \cmidrule(lr){3-4} \cmidrule(lr){5-6} \cmidrule(lr){8-13}
         &   &  I\$   &  Rest  &  Dec   &  Rest &    &  L/S  &  VM  & Mult  &  ALU  &  D\$   &  Rest   &    &   &   &   &  \\
        \midrule
        Mul & 0.30 & 4.72 & 0.51 & 0.01 & 0.09 & 1.42 & 0.22 & 3.46 & 0.97 & 0.02 & 5.53 & 0.07 & 0.05 & 0.22 & 4.25 & 0.76 & 22.60 \\
        \% & 1.33 & 20.88 & 2.26 & 0.04 & 0.40 & 6.28 & 0.97 & 15.31 & 4.29 & 0.09 & 24.47 & 0.31 & 0.22 & 0.97 & 18.81 & 3.36 & 100.00 \\[0.1cm]
        Div & 0.25 & 3.19 & 0.35 & 0.00 & 0.02 & 1.11 & 0.22 & 3.43 & 0.68 & 0.00 & 5.54 & 0.05 & 0.02 & 0.20 & 4.07 & 0.81 & 19.94 \\
        \%  & 1.25 & 16.00 & 1.76 & 0.00 & 0.10 & 5.57 & 1.10 & 17.20 & 3.41 & 0.00 & 27.78 & 0.25 & 0.10 & 1.00 & 20.41 & 4.06 & 100.00 \\[0.1cm]
        LS w/ VM & 0.32 & 4.63 & 0.54 & 0.01 & 0.09 & 1.38 & 0.30 & 3.50 & 0.09 & 0.03 & 9.18 & 0.18 & 0.06 & 0.22 & 4.06 & 0.62 & 25.21 \\
        \% & 1.27 & 18.37 & 2.14 & 0.04 & 0.36 & 5.47 & 1.19 & 13.88 & 0.36 & 0.12 & 36.41 & 0.71 & 0.24 & 0.87 & 16.10 & 2.46 & 100.00 \\[0.1cm]
        LS w/o VM & 0.30 & 4.39 & 0.51 & 0.00 & 0.07 & 1.36 & 0.30 & 3.48 & 0.07 & 0.02 & 9.12 & 0.17 & 0.06 & 0.22 & 4.04 & 0.64 & 24.75 \\
        \% & 1.21 & 17.74 & 2.06 & 0.00 & 0.28 & 5.49 & 1.21 & 14.06 & 0.28 & 0.08 & 36.85 & 0.69 & 0.24 & 0.89 & 16.32 & 2.59 & 100.00 \\[0.1cm]
        ALU & 0.30 & 4.36 & 0.50 & 0.05 & 0.13 & 1.69 & 0.24 & 3.47 & 0.11 & 0.03 & 5.53 & 0.08 & 0.08 & 0.24 & 4.05 & 0.72 & 21.58 \\
        \& & 1.39 & 20.20 & 2.32 & 0.23 & 0.60 & 7.83 & 1.11 & 16.08 & 0.51 & 0.14 & 25.63 & 0.37 & 0.37 & 1.11 & 18.77 & 3.34 & 100.00 \\[0.1cm]
        IGEMM & 0.61 & 10.17 & 1.59 & 0.19 & 0.65 & 5.88 & 0.61 & 3.84 & 4.41 & 0.71 & 13.75 & 1.00 & 0.31 & 1.12 & 4.68 & 2.28 & 51.80 \\
        \% & 1.18 & 19.63 & 3.07 & 0.37 & 1.25 & 11.35 & 1.18 & 7.41 & 8.51 & 1.37 & 26.54 & 1.93 & 0.60 & 2.16 & 9.03 & 4.40 & 100.00 \\ \midrule
        Leakage & 0.02 & 0.11 & 0.02 & 0.00 & 0.00 & 0.12 & 0.02 & 0.07 & 0.08 & 0.01 & 0.33 & 0.04 & 0.01 & 0.05 & 0.00 & 0.20 & 1.08 \\
    \bottomrule
    \end{tabular}
\end{table*}

\begin{figure}
    \includegraphics[width=\linewidth]{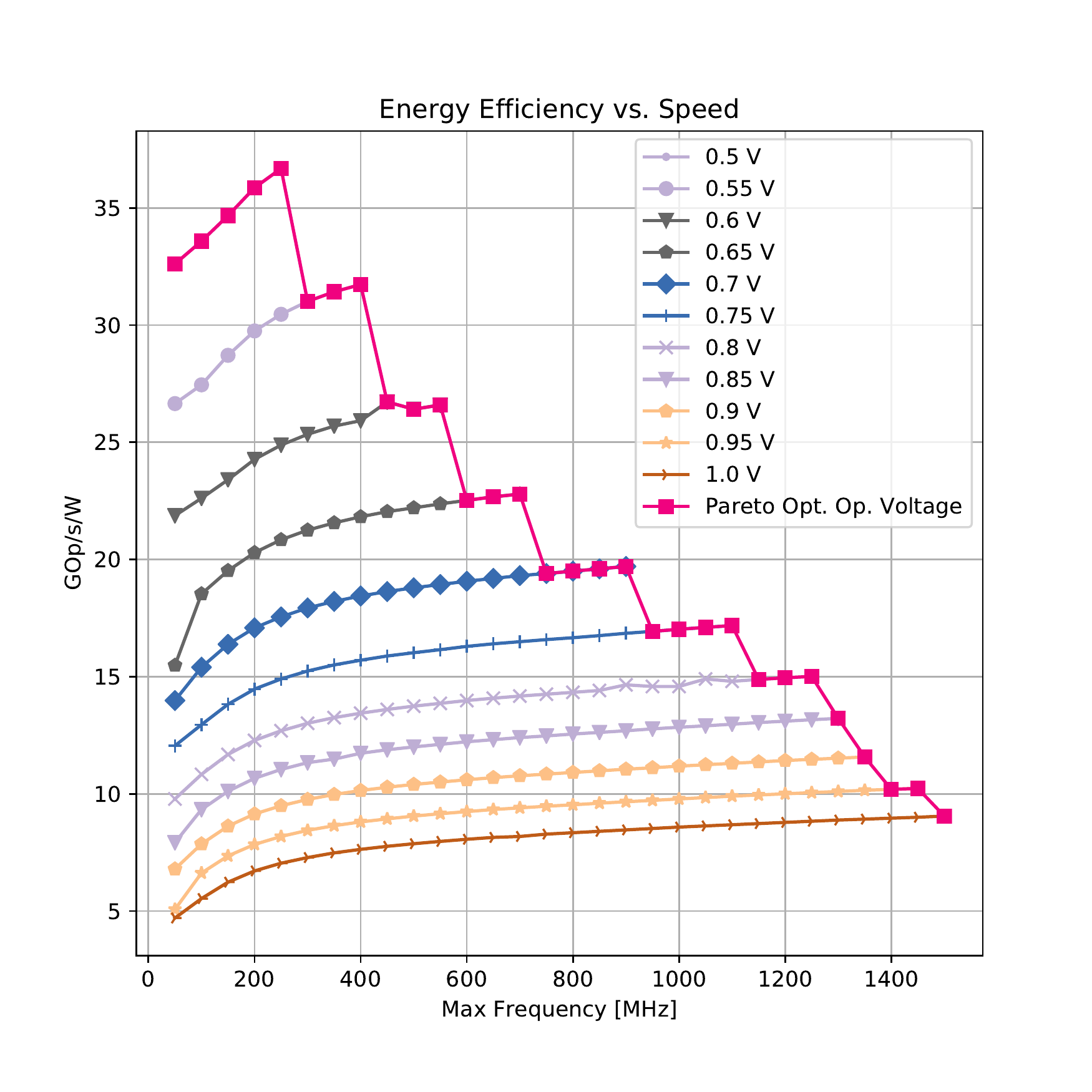}
    \label{fig:power_measurements}
    \caption{Detailed power measurement for mixed instruction workload (IGEMM) at different operating points (voltage and frequency). As the maximum speed is determined by the operating voltage we have indicated the most efficient operating point as a Pareto front on the right. Efficiency decreases for slower frequencies as leakage starts to dominate the power consumption. See Section~\ref{subsec:discussion} for a detailed discussion about power and energy efficiency.}
\end{figure}

\begin{figure}
    \includegraphics[width=\linewidth]{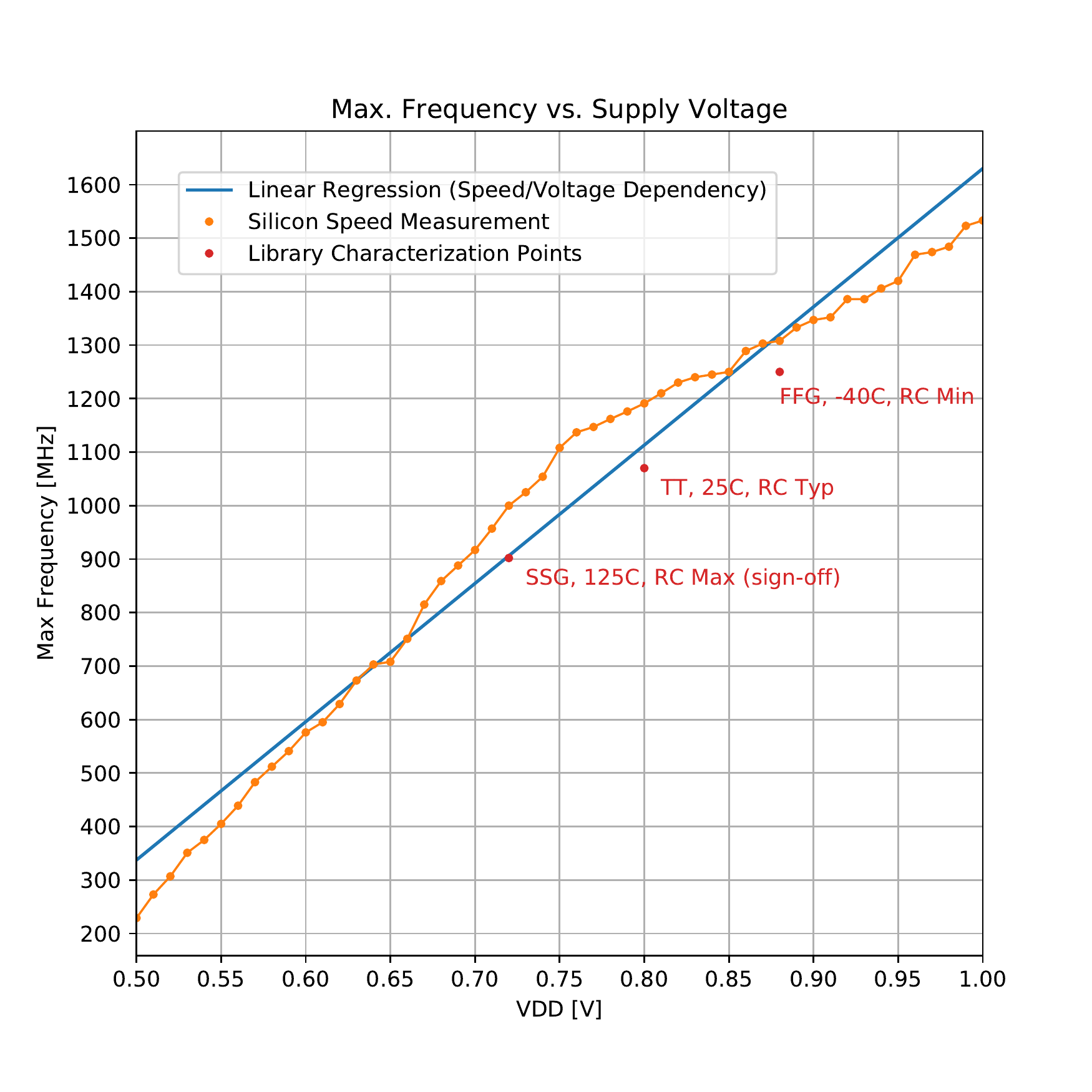}
    \caption{Frequency measurements under different supply voltages with zero Forward Body Bias (FBB). Worst, typical and best case library characterization points are denoted in red. Sign-off was done at \SI{902}{\mega\hertz} worst case. The linear regression fits the linear dependency between VDD and frequency. Section~\ref{subsubsec:caches} gives further details on worst case paths and achievable frequency.}
    \label{fig:max_freq}
\end{figure}

\begin{figure}
    \includegraphics[width=\linewidth]{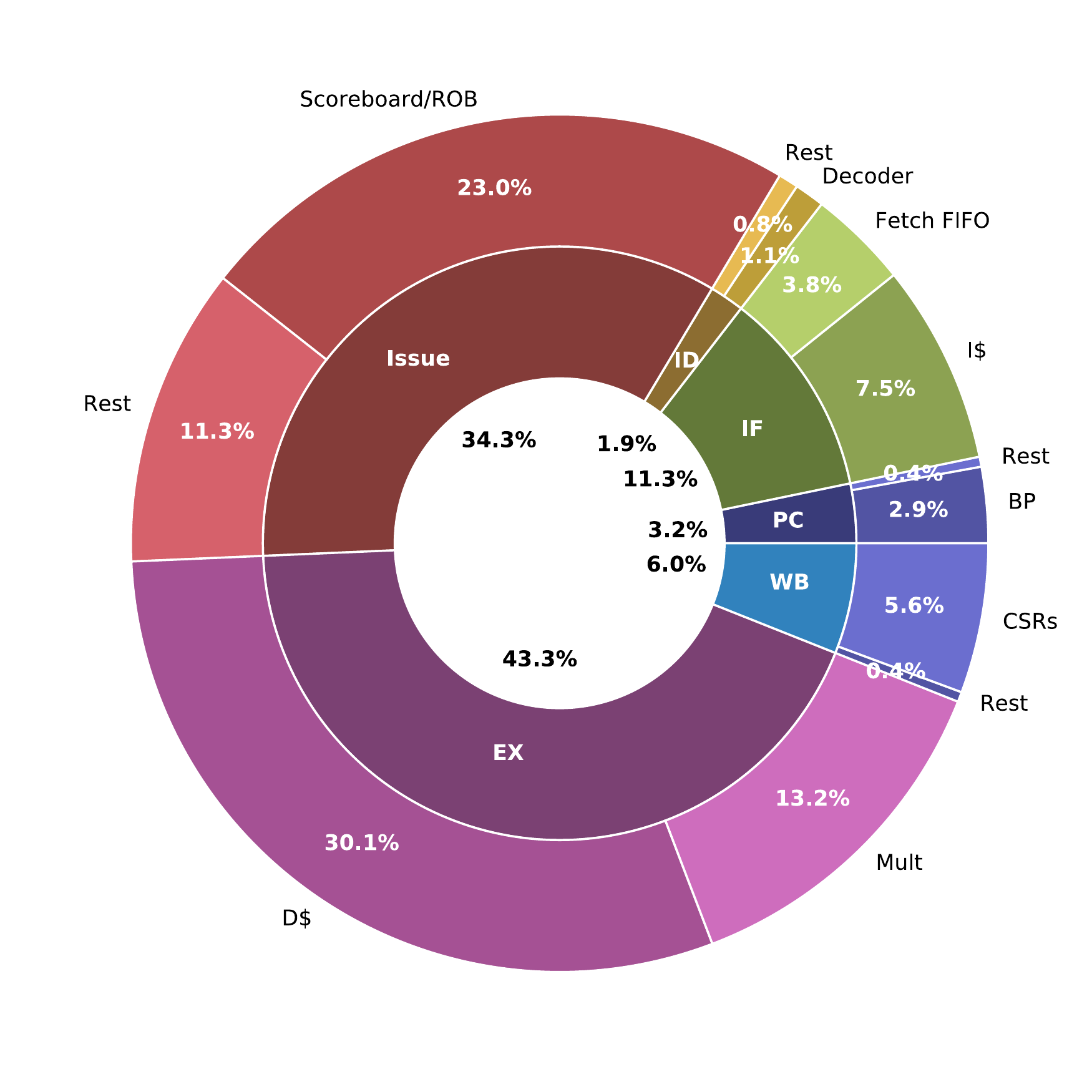}
    \caption{Detailed Area Breakdown. The total core area without cache memories amounts to \SI{210}{\kilo GE} @ 1.5 ns}
    \label{fig:area}
\end{figure}

\begin{figure}
    \includegraphics[width=\linewidth]{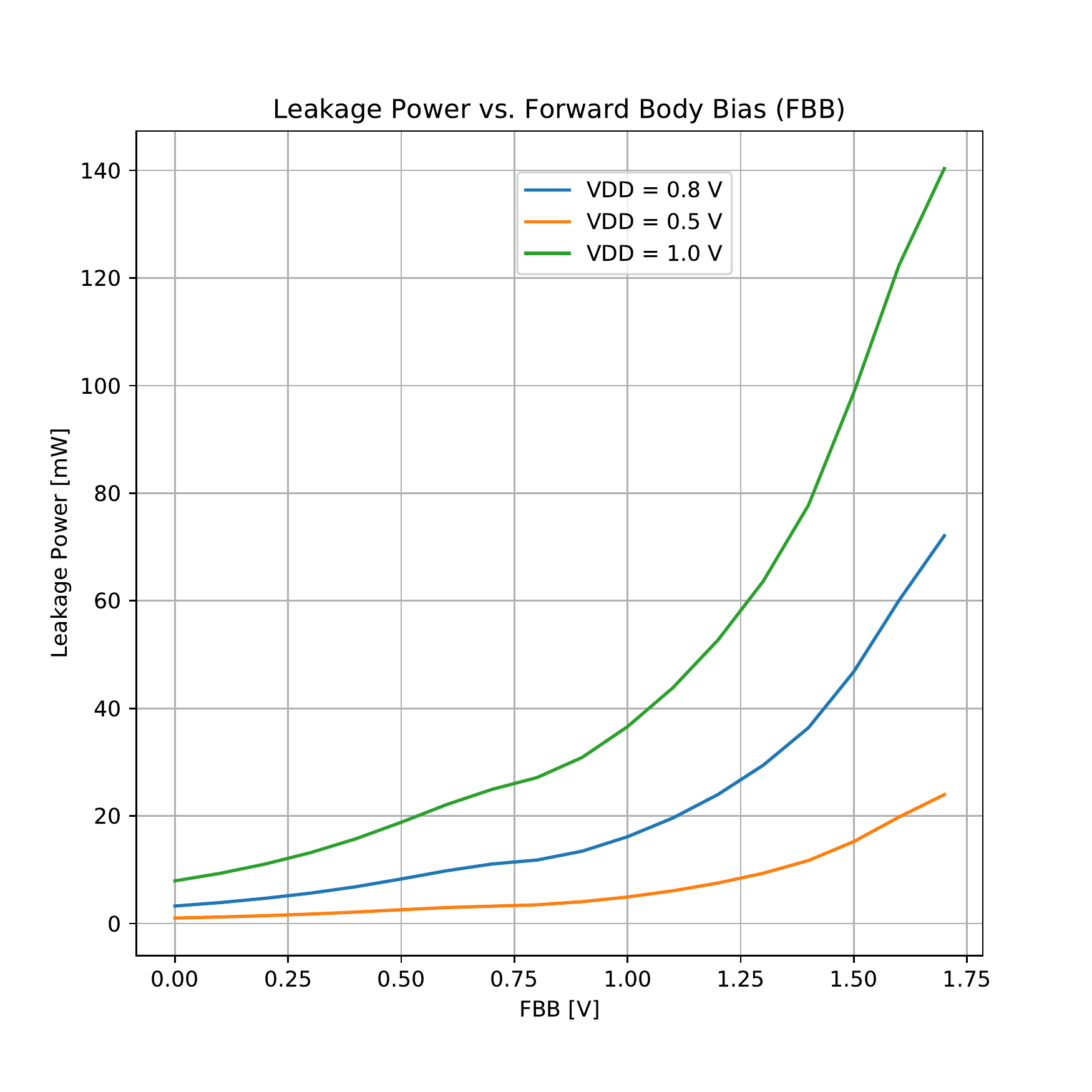}
    \caption{Leakage power for different FBB voltages and supply voltage at 0.5. 0.8 and \SI{1.0}{\volt}.}
    \label{fig:leakage}
\end{figure}

We measured our silicon implementation using an \textsc{Advantest 93000} industry-grade ASIC tester. Post-layout power numbers have been obtained using the post-layout floorplan and netlist of the fabricated design.

\subsection{Methodology}

We have developed a number of assembly-level tests which exercise particular architectural elements to provide a classification for the different instruction groups and hardware modules commonly found in the RISC-V \gls{isa} manual. In particular we focused on the following

\begin{itemize}
    \item \textbf{\gls{alu} instructions}: The \gls{alu} is used in the majority of RISC-V base instructions. It is used to calculate branch outcomes, arithmetic and logic operations. 
    \item \textbf{Multiplications}: The multiplier is a fully pipelined 2-cycle multiplier, hence it is a relatively big design and consumes considerable amounts of power. 
    \item \textbf{Divisions}: The divide algorithm is a radix-2 iterative divider. Hence area and power overhead are small but divisions can take up to 64 cycles. An early out mechanism can reduce division time significantly (depending on the operands). The test exercises long divisions.
    \item \textbf{Load and stores without virtual memory enabled}: For this test load and stores are triggered subsequently. The cache is warmed-up in this scenario. Address translation is not activated and the program operates in machine mode.
    \item \textbf{Load and stores with virtual memory enabled}: Similar to the above program except that the program is run in supervisor mode with address translation enabled. Both \gls{tlb}s are regularly flushed to trigger a page fault and activate the \gls{ptw}.
    \item \textbf{Mixed workload}: This s a generalized matrix-matrix multiplication. This test provides a compute intensive real-world workload triggering all architectural features. Furthermore this test is used for speed measurements. 
\end{itemize}

The tests have been run on silicon and on the post-layout netlist to provide a detailed per-unit listing. Separate power supplies for the core area, cache memory array and periphery allow for detailed power measurements on the actual silicon.
Post-layout and silicon measurements lie within a 10 \% error margin and a calibration factor has been applied to the post-layout power estimates to be aligned with our silicon measurements. We provide detailed results in Table~\ref{tab:energy_per_op}, separately listing the major contributors to power dissipation.

\subsection{Discussion}
\label{subsec:discussion}

We report up to 1.65 DMIPS/MHz for Ariane depending on the branch-prediction configuration and load latency (number of registers after the data cache). On the rather small Dhrystone benchmark the mispredict rate is 5.77\% with a 128-entry \gls{bht} and a 64-entry \gls{btb}. This results in an \gls{ipc} of 0.82 for the Dhrystone benchmark.

\subsubsection{Instruction and data caches}
\label{subsubsec:caches}
As can be seen in the area overview (Figure~\ref{fig:area}) and on the floorplan (Figure~\ref{fig:floorplan}) the largest units are the private L1 instruction and data caches. Furthermore, the critical path in the design is from the memories as the propagation delay of the slower SRAMs adds to the already costly 8-way tag comparison and data selection process. In addition, the wire delay and the diminishing routing resources close to and over the SRAM macros makes routing challenging.

Figure~\ref{fig:max_freq} plots the maximum frequency over a large selection of operating voltages as well as a fitted linear regression which captures the linear dependency between VDD and Fmax. The operating voltage is \SIrange{0.5}{1.15}{\volt} with a frequency from \SI{220}{\mega\hertz} to \SI{1.7}{\giga\hertz}. Below \SI{0.5}{\volt} the cache SRAMs are no longer functional. Body-biasing (BB) provides an additional tuning knob which allows for trading more power for higher frequency. The chip allows for forward body-bias (FBB) from \SI{0.0}{\volt} to \SI{0.5}{\volt} which results in up to 30\% higher speed at \SI{0.5}{\volt}. This results confirms that body biasing is very effective at low supply voltage to compensate for PVT variations, and to give a significant speed boost for run-to-halt operation. On the other hand, at higher voltages, the achievable speed-up reduces (only 6\% at \SI{1}{\volt}) and leakage power dissipation increases exponentially (see Figure~\ref{fig:leakage}). 

\subsubsection{Application-class features}
The impact of virtual memory (including \gls{tlb} and \gls{ptw}) is significant on the overall power dissipation. In particular on every instruction and data access a lookup in the \gls{tlb} has to be performed which can account for up to 27\% of the overall instruction energy while the energy used for actual computation is below 1\% in the case of \gls{alu}-centric instructions.

Furthermore the support for user and supervisor mode, and other architectural state like programmable interrupt vector and status registers, adds a significant area and power overhead on the \gls{csr} file. The difference between virtual memory activated and deactivated is quite small as the largest impact on power dissipation is the parallel indexing in the fully-set-associative \gls{tlb}'s which is also performed when address translation is disabled. The main difference is the regular page-table walking which does not have a large impact as the \gls{ptw} is not on the critical path and iteratively walks the tables.

\subsubsection{Multiplication and divisions}

The multiplier consumes over 13\% of the overall core area and can consume up to \SI{4.4}{\pico\joule} per cycle when active. The serial division unit was optimized for area and energy efficiency hence its overall impact is minimal.

\subsection{Comparison with non-application-class cores}

In comparison with a smaller, 32\,bit, embedded profile RISC-V core as implemented and reported in \cite{pullini2018mr}, we can see a considerable overhead mostly associated with the increased bit width, L1 caches and the application-class profile of Ariane. They report \SI{12.5}{\pico\joule} in \SI{40}{\nano\meter} technology for an integer matrix multiplication, which is comparable to the workload we used and reported (cf. Table~\ref{tab:energy_per_op}). Adjusting for technology scaling gains from \SI{40}{\nano\meter} to \SI{22}{\nano\meter} \cite{dreslinski2010near}  we compare \SI{10}{\pico\joule} of the small core to \SI{51.8}{\pico\joule} of Ariane. The same 32\,bit core has also been manufactured in \textsc{Globalfoundries 22 FDX} on the same die as part of Ariane's SoC. The small core achieves \SI{12.5}{\pico\joule} per instruction at \SI{0.8}{\volt}~\cite{schiavone2018quentin}. Considering that there is a power overhead involved with the larger SRAMs, Standard Cell Memories (SCM) and supporting logic which are also part of the \textsc{22\,FDX} design results, the estimated energy per instruction is comparable to our technology scaled estimate of \SI{10}{\pico\joule}.

Most of the overhead stems from the private L1 caches used in Ariane. The memory macros are power hungry and impose significant challenges during physical design. Another contributor to increased power consumption is the larger bit width. The architectural state (register file and \gls{csr} file) effectively doubles. This accounts for a 5.7\,\% (12\,kGE) area overhead for the register file and 2.8\,\% (6\,kGE) area overhead for the \gls{csr} file. Resulting in larger leakage power and increased switching power, both in the clock tree and on the registers themselves. Furthermore, also the datapath (e.g.: the functional units like \gls{alu} and multiplier) suffers from increased complexity, more logic area and tighter timing requirements resulting in decreased energy efficiency.
Last but not least, the overhead associated with the support for virtual memory is non negligible. \gls{tlb}s and \gls{ptw} are consuming up to \SI{3.8}{\pico\joule} per instruction which is a significant 38\% of the whole 32\,bit core.  

Loss of \gls{ipc} mainly results from mis-predicted branches and load data dependencies which need to stall subsequent, dependent instructions because of the three cycles latency of the data cache. Branch prediction can be improved by using more sophisticated prediction schemes like gshare, loop predictors or tournament predictors. However, branch prediction is a well researched topic~\cite{smith1981study} and has not been explored in this work. The load latency can be decreased at the expense of increased cycle time.

Since the speed of Ariane is higher than the speed reported for the embedded profile core, and its \gls{ipc} is comparable \cite{gautschi2017near}, we conclude that execution time is lower, although less energy efficient when comparing only the baseline RISCV \gls{isa}. However, the \gls{isa} extensions proposed in \cite{gautschi2017near} such as post-incrementing load and stores, hardware loops and \gls{simd} capability show a speedup of up to 10x compared to the baseline \gls{isa}. Hence, greatly outperforming pure architectural hardware performance enhancements (like scoreboarding and branch predicition) both in execution time and energy-efficiency. 

\begin{table}[!t]
    \renewcommand{\arraystretch}{1.3}
        \caption{16\,bit 2D (5x5) Convolution Benchmark}
        \label{tab:execution_time}
        \centering
        \begin{tabular}{@{}lrrrr@{}}
        \toprule
        & \bfseries Ariane & \multicolumn{3}{c}{\bfseries RI5CY~\cite{gautschi2017near}} \\
        \cmidrule(lr){2-2} \cmidrule(lr){3-5}
        ISA & RV64 & RV32 & RV32 + DSP & RV32 + SIMD \\
        \midrule
        Instructions [$\times10^3$] & 129 & 135 & 110 & 29 \\
        Cycles [$\times10^3$] & 152 & 137 & 117 & 31 \\
        IPC & 0.85 & 0.99 & 0.94 & 0.93 \\
        Freq. [\SI{}{\mega\hertz}] & 1700 & 690 & 690 & 690 \\
        Ex. Time [\SI{}{\micro\second}] & 89.5 & 198.8 & 179.7 & 45.2 \\
    \bottomrule
    \end{tabular}
\end{table}

Table~\ref{tab:execution_time} quantifies the above observation through an example of a compute-bound, 2D ($5\times5$ filter kernel) convolution of 16\,bit data types. The vanilla RISC-V baseline is 129\,k instructions for the 64\,bit \gls{isa} and 135\,k instructions for the 32\,bit \gls{isa}. When using the DSP \gls{isa} extensions the number of executed instructions drops to 110\,k instruction. At this stage all optimizations are automatically inferred by the compiler. Additional hand-tuning and usage of GCC's \gls{simd} builtins the executed instructions further reduces the instruction count to 29\,k instructions. While the \gls{ipc} is higher for the 32\,bit core, the higher clock frequency of Ariane significantly reduces the execution time. Nevertheless, all the proposed instruction set extensions reduce the amount of retired instructions by a factor of 4.6 compared to the 32\,bit RISC-V baseline, overall resulting in half the execution time compared to Ariane. We can therefore conclude that the cost for fundamental ``application-class'' microarchitectural features is significant, even within a simple, single-issue, in-order microarchitecture. Furthermore, we observe that computer performance and energy efficiency can be boosted more effectively with ISA extensions than with pure clock speed optimization. 

%% file: sec_relwork.tex
\section{Related Work} 
\label{sec:relwork}

Although RISC-V is a relatively young ISA there already exists a plethora of different commercial and open-source microarchitectural implementations. Currently most of them focus on the base integer subset and do not feature more sophisticated application-class features, like virtual memory.

To our knowledge, this is the first study on energy breakdown on the various functional units and on design rationale for the various features and links to ISA requirements on a competitive 64 bit, application-class core. 

One highly configurable application-class implementation of the RISC-V ISA is the Rocketchip generator. In particular the Rocketchip generator is a SoC generator, capable of generating highly parameterized cache-coherent multi-core systems \cite{asanovic2016rocket}, written in Chisel -- a hardware DSL embedded in Scala. The default Rocket core is a 5-stage, in-order core which can be parameterized to be either 64 or 32 bit. The core achieves up to \SI{1.3}{\giga\hertz} in \SI{45}{\nano\meter}~\cite{lee201445nm}. They report 1.72 DMIPS/MHz~\cite{lee2015raven}. The architecture is comparable to Ariane. They achieve a slightly higher \gls{ipc} at the expense of a slower clock speed. They report a core power efficiency of \SI{34}{\pico\joule} in \SI{40}{\nano\meter} TSMC technology. Considering technology scaling this results in \SI{27.2}{\pico\joule} per instruction in our target technology. When taking the cache memories into account (as reported in Table~\ref{tab:energy_per_op}) this results in an efficiency of \SI{46.8}{\pico\joule} per instruction, which is comparable to \SI{51.8}{\pico\joule} which we report. 

In the Rocketchip generator the single-issue, in-order core can be swapped with a super-scalar, out-of-order core called Boom \cite{Celio:EECS-2017-157,asanovic2015berkeley}. They report an \gls{ipc} of 1.5 for a four-issue, out-of-order implementation at a frequency of \SI{1.5}{\giga\hertz} in \texttt{TSMC 45\,nm} technology. The increase in \gls{ipc} comes at the expense of a significantly higher hardware complexity of 590\,kGE\footnote{Estimated from a similar cell library available to the authors}. Unfortunately no power results have been published about Boom.

Another major effort to provide a variety of different RISC-V cores is the SHAKTI project \cite{gala2016shakti} led by IIT-Madras. One particular implementation which is similar to Ariane is the 64 bit, 5-stage, in-order C-class core. It has been fabricated in Intel \SI{22}{\nano\meter} FinFet technology and consumes about 90-\SI{100}{\milli\watt} and requires about \SI{175}{\kilo G}. It also targets mid-range compute systems from 200-\SI{800}{\mega\hertz}. They report 1.68 DMIPS/MHz~\cite{shakticlass}. Therefore the DMIPS/MHz is similar to Ariane but Ariane is running at approximately double the speed of the C-class core and consumes less power at higher speeds. Moreover, no detailed energy breakdown analysis has been published on SHAKTI.

As remarked earlier, no detailed energy efficiency analysis has been performed on these cores on a functional unit level and we present first results of this kind.

Many studies, e.g. \cite{azizi2010energy,li2006cmp} have researched the energy efficiency of processors through high-level design space exploration. However, most of these studies either do not have silicon calibrated analysis or do not go into analysis of the various contributions to energy and cost, mainly because most processors are commercial and have a closed (secret) microarchitecture~\cite{tam2018skylake}. Other studies have solely focused on analyzing and improving certain aspects of the microarchitecture like for example the register file~\cite{zeng2015design}.

Most of these studies on energy-efficient processors were based on proprietary \glspl{isa} and/or microarchitectures. This is also one of the key novelties of our work on Ariane: not only the \gls{isa} is open, but also the whole microarchitecture and RTL design. Hence, it is possible to reproduce our results independently, as well as using Ariane as a basis to modify and improve the microarchitecture and its implementation. 

%% file: sec_conc.tex

\section{Conclusion}
\label{sec:conc}
We have presented Ariane, a 64\,bit, single-issue, in-order core taped-out in 22\,nm FDSOI technology. Based on this microarchitecture, we provide a rigorous efficiency analysis of the RISC-V \gls{isa} and its hardware impact. 

Furthermore, Ariane has been open-sourced in February 2018 and is available for download on GitHub with a very liberal license for the industry and research community. We provide support for Verilator- and QuestaSim-based RTL simulation as well as a ready-to-go FPGA bitstream and a pre-built Linux image\footnote{https://github.com/pulp-platform/ariane/releases}.


Our analysis reveals that, although many of Ariane's complex features are necessary to run full-featured \gls{os}es, most of the computation can be done on simpler, non-application-class cores as they share the same base \gls{isa} but lack features such as address translation and different privilege levels. Future work should focus on ISA-heterogeneous systems with microarchitectures consisting of many compute-centric, bare-metal cores and only a few higher-performance application-class management cores, as proposed by early high-level architectural studies~\cite{greenhalgh2011big,liang2013wimpy}. We expect such systems to achieve a high gain in energy-efficiency while keeping the programming model reasonable by just using the highly-efficient embedded cores for unprivileged compute tasks. 

In contrast to licensed \gls{isa}s like ARM or x86, the openness of the RISC-V \gls{isa} and the availability of encoding space makes it possible to innovate and explore different architectures and \gls{isa} extensions to enhance the efficiency of future computing systems.